\documentclass[preprint,12pt]{elsarticle}

\usepackage{graphicx}

\usepackage{amssymb}
\usepackage{multirow}

\begin{document}

\begin{frontmatter}



\title{Coexistence of periods in a bisecting bifurcation}


\author[UV,IFIC]{V. Botella-Soler}
\author[UV]{J.A. Oteo}
\author[UV,IFIC]{J. Ros}

\address[UV]{Departament de F\'{\i}sica Te\`{o}rica, Universitat de
Val\`{e}ncia,  46100-Burjassot, Val\`{e}ncia, Spain}
\address[IFIC]{IFIC, Universitat de Val\`{e}ncia-CSIC, 46100-Burjassot, Val\`{e}ncia,
Spain}

\begin{abstract}
The inner structure of the attractor appearing when the
Varley-Gradwell-Hassell population model bifurcates from
regular to chaotic behaviour is studied. By algebraic and
geometric arguments the coexistence of a continuum of neutrally
stable limit cycles with different periods in the attractor is
explained.
\end{abstract}

\begin{keyword}
piecewise map\sep bifurcations\sep route to chaos


\end{keyword}

\end{frontmatter}


\section{Introduction}

Since the first discoveries of chaotic behavior in discrete
maps, there has been great interest in the description of how
the dynamics of a map changes from regular to chaotic, and
viceversa, when a parameter of the map is varied. The so-called
\textit{routes to chaos} have been widely studied for smooth
maps where well-known phenomena such as period-doubling,
saddle-node bifurcations or intermittency appear
\cite{sprott2003chaos, ott2002chaos, strogatz2000nonlinear}.
The study of piecewise defined maps, which are useful in the
mathematical description of mechanical systems with friction or
collisions and of electrical circuits with switching
components, has introduced new phenomenology
\cite{di2008piecewise,zhusubaliyev2003bifurcations}.

 Here we consider  a one-dimensional
piecewise smooth discontinuous map originally proposed in
studies on population ecology
\cite{varley1973insect,may1976bifurcations}, the
Varley-Gradwell-Hassell map (henceforth referred to as VGH
map). It is a three-parameter map composed of a linear part and
a power-law decreasing piece. It may present both regular and
chaotic behavior, and for a critical value of one of its
parameters the system undergoes an abrupt order-to-chaos
transition. In \cite{botella2009dynamics} it was shown that at
this transition point, the system has a continuum of neutrally
stable limit cycles, i.e. cycles with multiplier unity. That
situation appears also in two-dimensional maps and in a recent
work \cite{lozi} we have named such bifurcations
\textit{bisecting bifurcations} for a reason that will become
clear in the following sections. The interesting characteristic
of the bisecting bifurcation in the VGH map is that neutrally
stable limit cycles with different periods coexist at the
transition point. Their periods and their distribution in phase
space depend on the location of the discontinuity of the map.
Our previous results are here sharpened by invoking algebraic
and geometric arguments.

The structure of the paper is as follows. In Section
\ref{sec:VGH} we introduce the VGH map and its bisecting
bifurcation. In Section \ref{sec:Localization} we explain a way
to study algebraically the existence and location of this type
of bifurcations. In Section \ref{sec:Gexp}, using the cobweb
diagram of the map we provide a geometrical explanation of the
bisecting bifurcation as well as the reason for the coexistence
of different periods. In Section \ref{sec:structure} we detail
the structure of the attractor and of the basins of attraction
at the bifurcation point. And finally, in Section
\ref{sec:discussion} we discuss the main results presented in
the paper.

\section{The VGH map}\label{sec:VGH}

The VGH map is defined in $[0,\infty)$ and reads
\begin{equation} \label{eq:Varley1}
V(x)=\left\{
\begin{tabular}{ll}
      $rx,$ & $x\le c$, \\
      $rx^{1-b},$ & $x>c$,
     \end{tabular}
     \right.
\end{equation}
where $r,b>1$ and $c>0$. When $c=1$ the system is continuous,
being discontinuous otherwise. The VGH map presents both
regular and chaotic behavior and its dynamics exhibits a number
of phenomena including interior crises and crisis-induced
intermittency \cite{botella2009dynamics}.

\begin{figure}[H]
\includegraphics[scale=0.35]{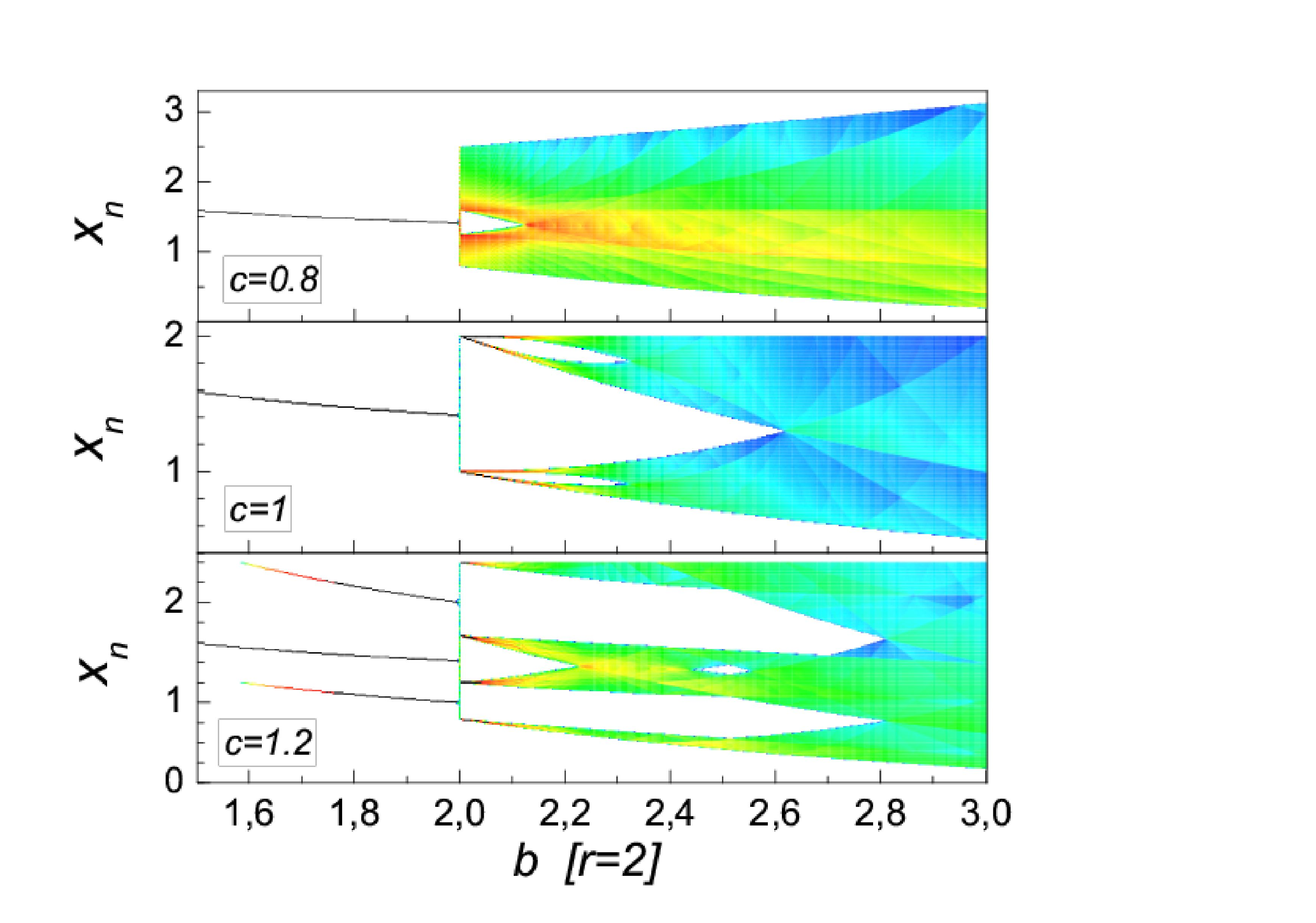}
\caption{\label{fig:VGHMap}Bifurcation diagrams of the VGH map
$V(x)$ with $r=2$ and three different values of $c$.}
\end{figure}

A customary construction of the bifurcation diagram by varying
$b$, for fixed $r$ and $c$, shows an order-to-chaos bifurcation
at $b=2$. For $b<2$ the system is regular with a single
periodic attractor whose period depends on $c$. When $b>2$ the
system is chaotic without further regular window. This is clear
from the bifurcation diagrams in Figure \ref{fig:VGHMap} where
three different values of the parameter c are considered. The
top and bottom panels of the figure stand for cases where the
map is discontinuous ($c=0.8, 1.2$) and the middle panel shows
the bifurcation diagram in the continuous case ($c=1$). At the
bifurcation point $b=2$ all these diagrams exhibit a vertical
segment which constitutes the focus of this work.

Detailed numerical experiments, buttressed by the arguments in
the next sections, demonstrate that any initial condition end
up in a neutrally stable limit cycle. There is precisely an
infinity of them which form the vertical segment in the
diagrams of Figure \ref{fig:VGHMap}. It is worthwhile to point
out that the numerical detection of this continuum of
attractors which mediates the bifurcation presents three main
difficulties. First of all it exists only for the precise
numerical value $b=2$. Then, an accurate enough
machine-representation of the parameter values must be used in
the sampling. Secondly, as the number of neutrally stable
periodic orbits that exist for $b=2$ is infinite, a large
enough number of initial conditions must be used to generate
the vertical segment in the diagram. Finally, the whole
attractor can easily be mistaken for a single chaotic attractor
unless properly scrutinized. The algebraic and geometric
analyses in the next sections are of great help in the correct
identification of the vertical segment in the bifurcation
diagram.

\section{Localization of the bifurcation in parameter
space}\label{sec:Localization}

In this section we prove that for the VGH map the bisecting
bifurcation takes place only at $b=2$. We start by introducing
a change of variables that permits us to exactly linearize the
VGH map
\begin{equation}
z\equiv 2\log(x)/\log(r), \quad \xi \equiv 2\log(c)/\log(r).
\end{equation}
In terms of $z$ and $\xi$, the VGH map can be cast in the
following linear form
\begin{equation}\label{eq:linVGH}
L(z)=\left\{
\begin{tabular}{ll}
      $z + 2,$ & $z\le \xi$, \\
      $(1-b)z +2,$ & $z>\xi$.
     \end{tabular}
     \right.
 \nonumber
\end{equation}
This linearized version allows the following algebraic
approach. In the new form $L(z)$, the elements $\{z_1,\, z_2,\,
\ldots,\, z_T \}$ of a cycle of period $T$ satisfy
\begin{eqnarray}
  z_n &=& \left\{\begin{tabular}{ll}
                 $z_{n-1}+2,$ & $z_{n-1}\leq \xi$,\\
                  $(1-b)z_{n-1}+2$ & $z_{n-1}> \xi$, \quad $n=2,\ldots,
                  T$,
                    \end{tabular}
                    \right.\nonumber
   \\
  z_1 &=& \left\{\begin{tabular}{ll}
                   $z_{T}+2$, &  $z_{T}\leq \xi$,   \\
                   $(1-b)z_{T}+2$, & $z_{T}> \xi$.
                   \end{tabular}
                   \right.
\end{eqnarray}
These expressions can be cast in the form
\begin{eqnarray}\label{eq:system}
 z_1+\alpha_{T}z_{T}&=&2,\nonumber\\
  z_n+\alpha_{n-1}z_{n-1}&=&2,\quad n=2,3,\ldots,T,
\end{eqnarray}
with coefficients
\begin{equation}
    \alpha_k=\left\{
\begin{tabular}{ll}
      $-1,$ & $z_k\le \xi$, \\
      $(b-1),$ & $z_k>\xi$.
     \end{tabular}
     \right.
\end{equation}
The linear system of equations (\ref{eq:system}) can be
expressed in matrix form
\begin{equation}
    \sum\limits_{j=1}^{T }A_{ij}z_j=2, \quad i=1,2,\ldots,
    T,
    \end{equation}
where the matrix $A$ is
\[
A=\left(
  \begin{array}{cccccc}
    \alpha_1 & 1 & 0 & \ldots & \ldots & 0 \\
    0 & \alpha_2 & 1 & 0 & \ldots & 0 \\
    0 & 0 & \alpha_3 & 1 & \ldots & 0 \\
    \ldots & \ldots & \ldots  & \ldots  & \ldots  & \ldots \\
    0 & 0 & \ldots  & \ldots  & \alpha_{T-1} & 1 \\
    1 & 0 & \ldots  & \ldots  & 0 & \alpha_T \\
  \end{array}
\right).
\]

In order to find an attractor consisting of infinitely many
limit cycles of period $T$ we need the system of equations
(\ref{eq:system}) to be compatible and indeterminate. A
necessary condition is therefore $\det(A)=0$. The determinant
of the matrix $A$ is
\begin{equation}
\det(A)=\prod_{k=1}^{T}\alpha_k\, +\, (-1)^{T+1}.
\end{equation}
If the cycle has $d$ of its elements satisfying $z_i>\xi$ this
gives
\begin{equation}
\det(A)=(-1)^{T-d}(b-1)^d +\, (-1)^{T+1},
\end{equation}
which vanishes  for $b=0$ and, if $d$ is even, also for $b=2$.
Since we are only interested in $b>1$, the only possible
solution is then $b=2$.  This is precisely the value found from
the bifurcation diagrams in Section \ref{sec:VGH}. It is
important to notice that this value of $b$ and equation
(\ref{eq:linVGH}) imply neutral stability for any possible
cycle. Of course, the argument in this section does not
complete the analysis of the vertical segment in those
diagrams. It only excludes any other value of $b$ as candidate.
In the next Section we will explicitly show that the infinite
set of periodic orbits actually exist.

\section{Geometrical explanation of the bisecting bifurcation}\label{sec:Gexp}

\begin{figure}[H]
\includegraphics[scale=0.3]{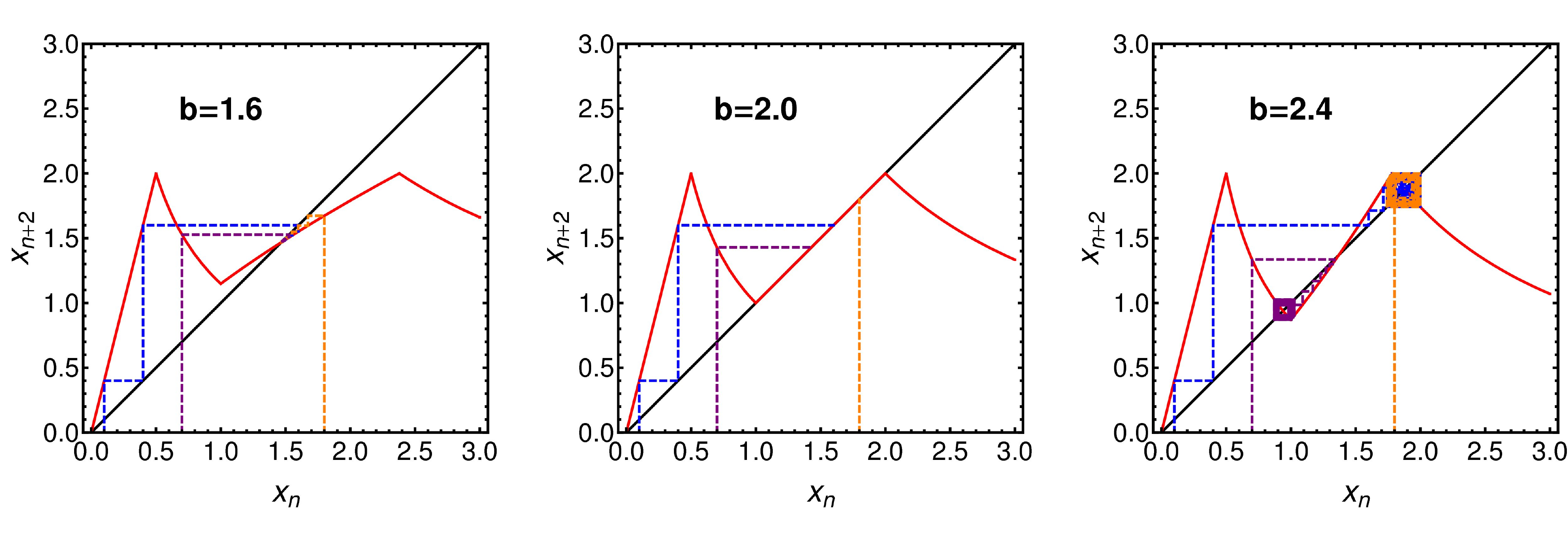}
\caption{\label{fig:Cobweb}A sequence of three cobweb plots of
$V^{[2]}(x)$ illustrating the bisecting bifurcation from order
to chaos at $b=2$. Here $c=1,r=2$. The presence of the segment
of fixed points is apparent in the case $b=2$.}
\end{figure}

Next we develop an heuristic explanation for the emergence of a
bisecting bifurcation based on the cobweb diagram of a general
one-dimensional piecewise defined map $f(x;p)$ with parameter
$p$. Notice that an infinite set of regular trajectories of
period $n$ will be observed for a critical value $p_c$ of the
parameter when  the $n$th iterate of the map $f^{[n]}$ has a
piece which is co-linear with the bisectrix. Hence, the name we
are using for these bifurcations.  More explicitly, at the
bifurcation,  $f^{[n]}$ will be of the form
\begin{equation} \label{eq:iter}
f^{[n]}(x;p_c)=\left\{
\begin{tabular}{ll}
      $\cdots ,$ & $\cdots $, \\
      $x,$ & $x\in(x_l,x_r)$,\\
      $\cdots ,$ & $\cdots $
     \end{tabular}
     \right.
\end{equation}
An instance of this phenomenon in the VGH map is given by its
second iterate which reads
\begin{equation} \label{eq:2iter}
V^{[2]}(x)=\left\{
\begin{tabular}{ll}
      $r^2x,$ & $x\le\frac{c}{r}$, \\
      $r^{2-b}x^{1-b},$ & $\frac{c}{r}<x\le c$,\\
      $r^{2-b}x^{(1-b)^2} ,$ & $c<x\le (\frac{r}{c})^{\frac{1}{b-1}}$,\\
      $r^2 x^{1-b} ,$ & $x>c, x>(\frac{r}{c})^{\frac{1}{b-1}}$.
     \end{tabular}
     \right.
\end{equation}
When $b=2$ the third piece becomes $x$ and we find a bisecting
bifurcation mediated by an infinite set of neutrally stable
period-2 limit cycles provided $c<\sqrt{r}$. This is
illustrated in Figure \ref{fig:Cobweb} where three cobweb plots
of $V^{[2]}(x)$ for values $b=1.6, 2.0, 2.4$ and $c=1$ are
shown. In the three cases the same three initial conditions
have been evaluated. When $b=1.6$, all trajectories converge to
the stable fixed point. In the critical case, when $b=2$, each
initial condition evolves to a different neutrally stable fixed
point of $V^{[2]}(x)$. For $b=2.4$ the trajectories are
chaotic.

\begin{figure}[H]
\includegraphics[scale=0.4]{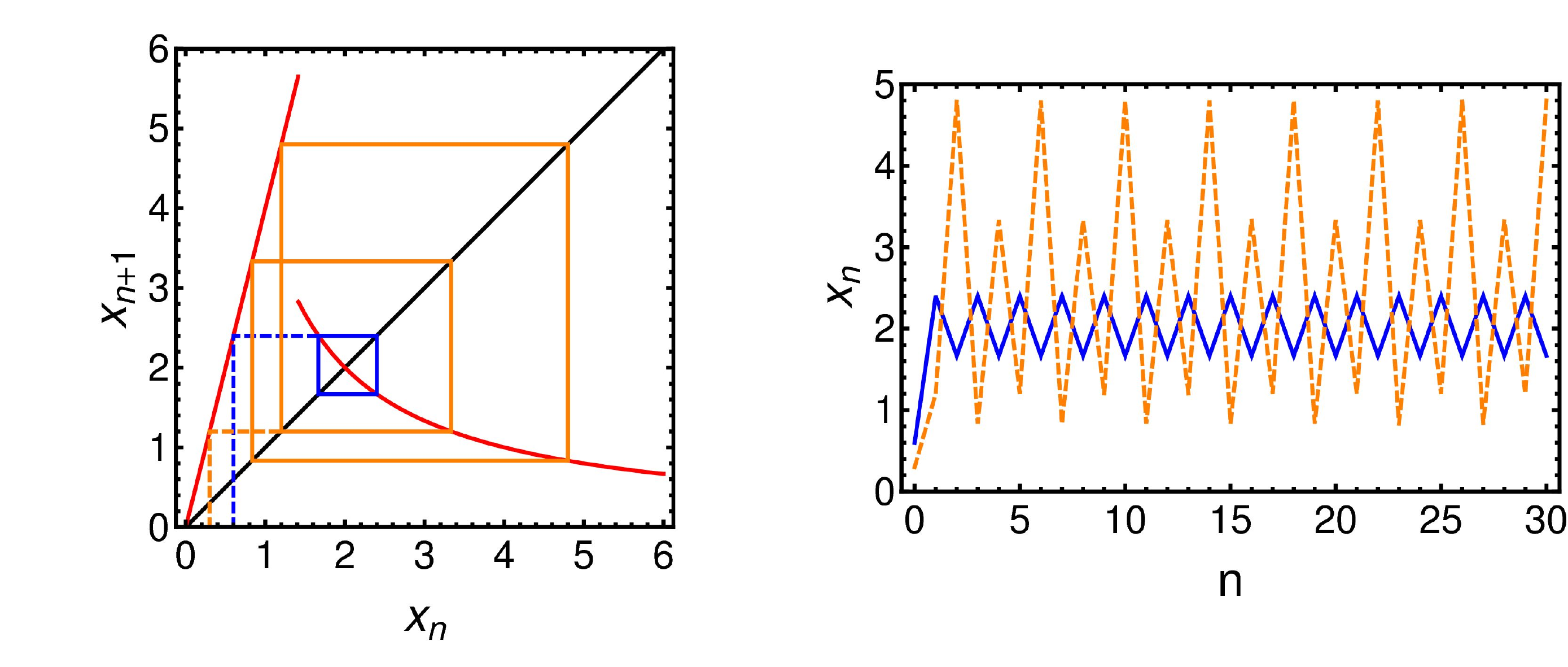}
\caption{\label{fig:CSqrt2}Cobweb diagram (left panel) and
trajectories (right panel) of $V(x)$ for $b=2$, $c=\sqrt{2}$,
$r=4$ and two different initial conditions ($x_0=0.3,0.6$). It
is clearly seen that each initial condition enters a neutrally
stable limit cycle of different period.}
\end{figure}

In principle, several iterates of $f$ can have pieces co-linear
with the bisectrix simultaneously for the same value of $p_c$.
In such cases, infinite sets of cycles of different periods
will coexist when $p=p_c$. This is the case for the VGH map.
When $b=2$ and $c>1$ the neutrally stable limit cycles can show
different periods. This is illustrated in Figure
\ref{fig:CSqrt2} where the evolution in the cobweb diagram and
the trajectories of two different initial conditions are shown
for $b=2$, $r=4$ and $c=\sqrt{2}$. One of the trajectories
enters a period-2 limit cycle while the other evolves to a
period-4 limit cycle. The distribution of the limit cycles of
different periods in phase space is detailed in the next
Section.

\section{Attractors and basins of attraction for $b=2$}\label{sec:structure}

In this Section we study at depth the structure of the
attracting segment of the VGH system at the critical point
$b=2$. To facilitate the analysis we express the VHG map for
$b=2$ in terms of the new variable $w\equiv z-1$ as
\begin{equation}
W(w)=\left\{
\begin{tabular}{ll}
      $w + 2,$ & $w\le \xi -1$, \\
      $-w ,$ & $w>\xi-1$.
     \end{tabular}
     \right.
 \nonumber
\end{equation}
One advantage of this form is that the segment of limit cycles,
$\mathcal{A}\equiv [-|\xi|-1,|\xi|+1]$, is symmetric in phase
space with respect to $w=0$.

\subsection{Structure of the attractor}

We start our study by distinguishing the cases of positive and
negative $\xi$.

\subsubsection{$\xi\le0$}

This case presents the simplest dynamics. The segment
$\mathcal{A}$ is composed of infinite period-2 limit cycles
around $w=0$, which is a fixed point.

\subsubsection{$\xi>0$}

\begin{figure}[H]
\includegraphics[scale=0.30]{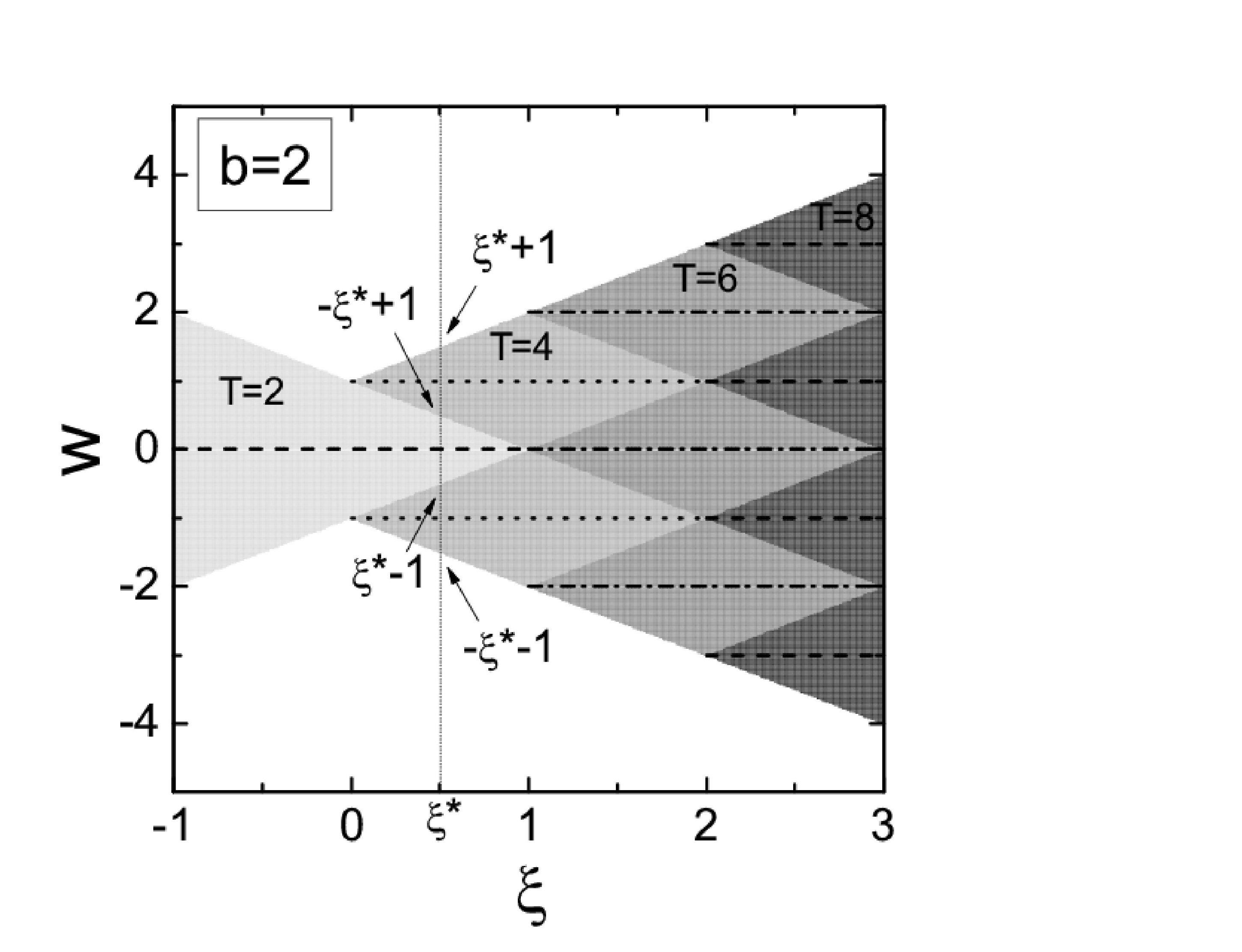}
\caption{\label{fig:Periodogram} Structure of the attracting
segment ($b=2$) in phase space for different values of $\xi$.
The different tones of gray code cycles with different periods.
The discontinuous horizontal lines stand for individual cycles
with integer elements of periods 1, 2, 3 and 4 (dashed, dotted,
dash-dotted and dashed respectively) embedded in the continuum
of cycles of the attracting segment.}
\end{figure}

The dynamics is more complicated for positive $\xi$. All
integer initial conditions $w_0\in\mathbb{Z}$ lead to limit
cycles with integer elements. In particular, if the initial
condition is even (resp. odd), the final limit cycle will have
as its elements all even (resp. odd) integers inside the
segment $\mathcal{A}$. The periods of these cycles, which
depend on $\xi$, are detailed in Table \ref{tab:T}. If
$w_0\notin\mathbb{Z}$ we need to study the cases
$\xi\in\mathbb{N}$ and $\xi\notin\mathbb{N}$ separately:

\begin{itemize}
\item{$\xi\notin\mathbb{N}$}

For positive non-integer $\xi$ we have two infinite sets of
limit cycles of periods $T=2(n+1)$ and $T=2(n+2)$ with
$n=\lfloor\xi\rfloor$ (where $\lfloor\cdot\rfloor$ stands for
the floor function). These cycles spread  over $\mathcal{A}$ in
a rather peculiar way. To make it clear we find useful to
consider the finite set of points
\begin{equation}
\mathcal{B} = \mathcal{A} \bigcap \{ w=(-1)^\alpha
\xi+(-1)^\beta(2k+1),\quad k=0\dots n,\quad\alpha,\beta=0,1 \}
\end{equation}
whose elements, when written in increasing order, we denote by
$w^i,\, i=1,\, 2, \ldots$. Then, $\mathcal{B}$ punctuates a
partition of the interval $\mathcal{A}$ in subintervals. Points
in the same subinterval belong to cycles with the same period.
Points in contiguous subintervals belong to cycles with
different period.

For instance, if $0<\xi<1$ then $n=0$ and the frontiers of the
subintervals are given by
\begin{equation}\label{eq:frontiers}
\mathcal{B}=\{w^1,w^2,w^3,w^4\}=\{-\xi-1,\xi-1,-\xi+1,\xi+1\}.
\end{equation}
Thus, in this case three subintervals exist inside the
attractor. In Table \ref{tab:T} we have detailed the period of
the limit cycle for initial conditions inside the attracting
segment $\mathcal{A}$.

\item{$\xi\in\mathbb{N}$}

When $\xi$ is a natural number every limit cycles has period
$T=2(\xi+1)$.

\end{itemize}
All this information is contained in Figure
\ref{fig:Periodogram} where we have plotted the structure of
the attracting segment at $b=2$ as a function of $\xi$ with
$\xi\in[-1,3]$. The different tones of gray stand for different
periods. As an example, a particular $\xi^{\ast}<1$ has been
chosen to illustrate the position of the frontiers given in
(\ref{eq:frontiers}). The discontinuous horizontal lines stand
for cycles with initial conditions $w_0\in\mathbb{Z}$.

\begin{table}[H]
\caption{\label{tab:T} Period of the trajectories according to
the initial point $w_0\in [-|\xi|-1,|\xi|+1]$, with $b=2$ and
$\xi > 0$ ($\xi\notin\mathbb{N}$). The description of the case
$w_0 \in \mathbb{Z}$ is also valid for $\xi\in\mathbb{N}$.}
\begin{center}
\begin{tabular}{lccr}
\hline
Initial point & $ $ & Period & Condition\\
\hline \multirow{2}{*}{$w_0 \in (w^i,w^{i+1})$ and $w_0 \notin
\mathbb{Z}$} &
\multirow{2}{*}{} & $2(n+1)$ &\mbox{even $i$}\\
 &  & $2(n+2)$&\mbox{odd $i$}\\
\hline $w_0=w^i$ and $w_0\notin\mathbb{Z}$ & & $2(n+2)$ & $\forall i$ \\
\hline \multirow{2}{*}{$w_0 \in \mathbb{Z}$} &
\mbox{ $N=\lfloor(\xi+1)/2\rfloor$} & $2N+1$ & even $w_0$  \\
 & \mbox{$M=\lfloor\xi/2\rfloor$ } & $2(M+1)$ & odd $w_0$  \\
\hline
\end{tabular}
\end{center}
\end{table}

\subsection{Basins of attraction}

\begin{figure}[H]
\includegraphics[scale=0.5]{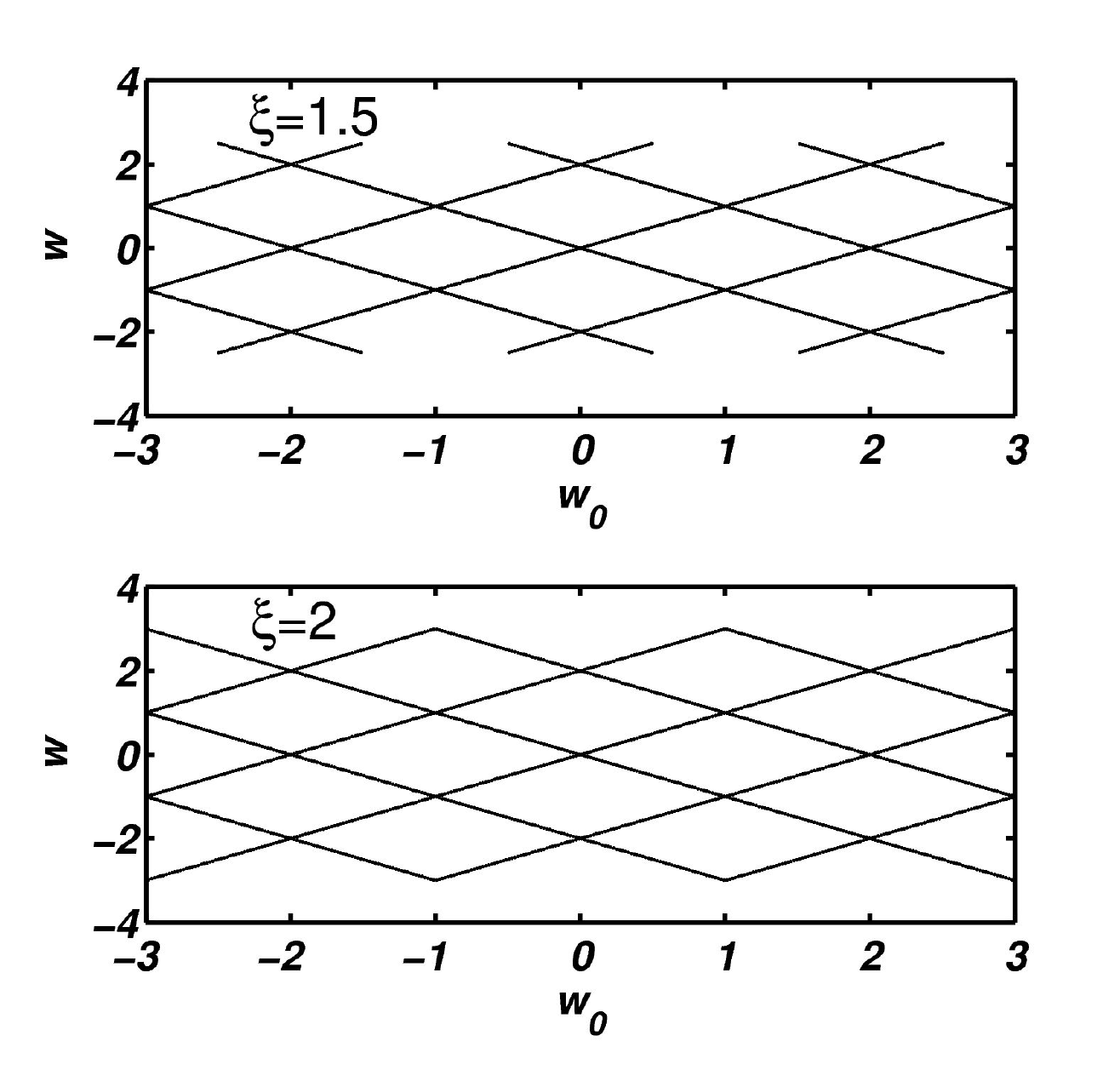}
\caption{\label{fig:Basins} Structure of the basins of
attraction for two different values of the discontinuity
location parameter $\xi$. The graphs show the final attractor
for each $w_0$. For $\xi=1.5$ (upper panel) infinite limit
cycles of periods 4 and 6 exist as well as the period-2 cycle
\{-1,1\} and the period-3 cycle \{-2,0,2\} can be seen. For
$\xi=2.0$ (lower panel) infinite limit cycles of period-6 are
present filling the space between the points of the period-3
cycle \{-2,0,2\} and the period-4 cycle \{-3,-1,1,3\}.}
\end{figure}

Given the limit cycle to which a point $w_0$ tends, its basin
of attraction can be written as
\begin{equation}\label{eq:basin}
B_{w_0}=\{w\in \mathbb{R}|w=\lfloor w_0\rfloor+2k\pm d,
k\in\mathbb{Z}\},\quad d=w_0-\lfloor w_0\rfloor.
\end{equation}
This structure of the basins of attraction is reflected in
Figure \ref{fig:Basins} for two different values of $\xi$. In
this figure for each initial condition $w_0\in [-3,3]$ we plot
the cycle in $\mathcal{A}$ which traps it. The figure suggests
a periodic structure in the horizontal direction. It reflects
the partition of phase space into equivalence classes
established by (\ref{eq:basin}). More specifically, all even,
odd and odd half-integer initial conditions constitute three
equivalence classes by themselves. Any other initial condition
generates its equivalence class by repeatedly adding
alternatively $2d$ and $2(1-d)$.

The combination of Table 1 and (\ref{eq:basin}) allows to
determine the period of the cycle to which an arbitrary point
$w_0$ tends. For the sake of illustration, consider the case
$\xi=1.5$ and the initial condition $w_0=2.7$. In this case
$n=\lfloor\xi\rfloor=1$ and the frontiers inside the segment
$\mathcal{A}$ are given by
\begin{equation}
\mathcal{B}=\{w^1,w^2,w^3,w^4,w^5,w^6\}=\{-2.5, -1.5, -0.5,
0.5, 1.5, 2.5\}.
\end{equation}
Since the chosen initial condition falls outside the attracting
segment we will make use of (\ref{eq:basin}) to determine
another initial condition $w_0'\in B_{w_0}$ leading to the same
final limit cycle. Choosing $k=0$ we can readily find
\begin{equation}
w_0'=\lfloor w_0\rfloor+2k-d=1.3 \in (w^4,w^5).
\end{equation}
If we now take into account the classification detailed in
Table \ref{tab:T} we can conclude that both $w_0$ and $w_0'$
will enter a limit cycle of period 4. It can be checked that
this is in fact the case in the upper panel of Figure
\ref{fig:Basins}.

\section{Discussion of the results}\label{sec:discussion}

This work focuses on the coexistence of different periods in
the set of neutrally stable limit cycles that mediates the
bisecting bifurcation of the VGH map. We have proved
algebraically that it occurs only for $b=2$. In principle, the
same method might be used to determine the existence and
position in parameter space of these bifurcations for any
piecewise linear map.

The geometric explanation complements the algebraic approach
and provides us with an intuitive way of interpreting the
bifurcation in terms of the cobweb diagram. In particular, it
allows us to understand the coexistence of different periods as
the simultaneous presence of pieces co-linear with the
bisectrix in different iterates of the map.

The structure of the attractor in $b=2$ has been described in
detail. Given a certain value of the discontinuity parameter
(either $c$ or $\xi$), the periods of the limit cycles and
their location in phase space can readily be determined. The
structure of the basins of attraction of the limit cycles has
been resolved as well, allowing us to associate each initial
condition to its final limit cycle.

The described bisecting bifurcation is accompanied by border
collisions
\cite{di2008piecewise,zhusubaliyev2003bifurcations,nusse1995border}
as can be seen in Figure \ref{fig:Cobweb}. In the example
described in that figure, two unstable fixed points of
$V^{[2]}(x)$ (for $b>2$) collide with discontinuities of the
map when $b=2$ and cease to exist for $b<2$. An attempt to
classify the phenomenology of border-collision bifurcations in
one-dimensional discontinuous maps can be found in
\cite{jain2003border}. The classification is based on the
linearization of the map around the collision point both in
phase space and parameter space. However, the bisecting
bifurcation presented here corresponds to one of the critical
cases explicitly excluded from this classification.

In this study we have restricted our attention to the VGH map
but what we have called bisecting bifurcations are present in
other continuous and discontinuous piecewise smooth maps.
However, to the best of our knowledge, they have very often
gone unnoticed in the literature. In the continuous case,
bisecting bifurcations can be observed in maps such as the skew
tent map \cite{nusse1998dynamics} or the map describing the
dynamics of the boost converter \cite{banerjee2000IEEE}. In
particular, in \cite{gardini2008growing} a continuous piecewise
smooth map introduced as a model of economic growth
\cite{matsuyama1999growing} is studied and the values of the
parameters for which a bisecting bifurcation takes place are
identified. Discontinuous maps candidates to show bisecting
bifurcations can be found in \cite{avrutin2006multi,
avrutin2008fully}. Our numerical experiments have shown this is
in fact the case. Moreover, the map studied in
\cite{avrutin2006multi} shows coexistence of different periods
in the set of neutrally stable limit cycles with a structure
very similar to the one described in this paper for the VGH
map.

\section*{Acknowledgements}
This work has been partially supported by contracts MCyT/FEDER,
Spain FIS2007-60133 and MICINN (AYA2010-22111-C03-02). VBS
thanks Generalitat Valenciana for financial support.





\bibliographystyle{elsarticle-num}
\bibliography{VGH2_ms}







\end{document}